\documentclass[ALICE,manyauthors]{cernphprep}

\usepackage[comma,square,numbers,sort&compress]{natbib}
\usepackage{hyperref}
\usepackage{lineno}
\usepackage{color}
\usepackage{multirow}

\newcommand{\pPb}{{p--Pb}\xspace}
\newcommand{\Pbp}{{Pb--p}\xspace}
\newcommand{\snn}{\ensuremath{\sqrt{s_{\mathrm{\scriptscriptstyle NN}}}}\xspace}
\newcommand{\jpsi}{\ensuremath{\mathrm{J}/\psi}\xspace}
\newcommand{\psip}{\ensuremath{\psi(\mathrm{2S})}\xspace}

\newcommand{\pt}{\ensuremath{p_{\mathrm{T}}}\xspace}
\newcommand{\Wgp}{\ensuremath{W_{\gamma \mathrm{p}}}\xspace}
\newcommand{\eXa}{\ensuremath{\epsilon\times\mathrm{A}}\xspace}

\begin{document}%

\begin{titlepage}
\PHyear{2018}
\PHnumber{236}      
\PHdate{28 August}  
%

\title{Energy dependence of exclusive  $\jpsi$ photoproduction off protons in ultra-peripheral \pPb collisions at $\snn = 5.02$  TeV}

\ShortTitle{Energy dependence of exclusive $\jpsi$ photoproduction ...}   

\Collaboration{ALICE Collaboration\thanks{See Appendix~\ref{app:collab} for the list of collaboration members}}
\ShortAuthor{ALICE Collaboration} 

\begin{abstract}
The ALICE Collaboration has measured the energy dependence of  exclusive photoproduction of \jpsi vector mesons off proton targets in ultra--peripheral \pPb collisions at a centre-of-mass energy per nucleon pair $\sqrt{s_{\rm NN}} = 5.02$ TeV. The e$^+$e$^-$ and $\mu^+\mu^-$ decay channels are used to measure the cross section as a function of the rapidity  of the \jpsi in the range $-2.5 < y < 2.7$, corresponding to an energy in the $\gamma$p centre-of-mass in the interval $40 < W_{\gamma\mathrm{p}}<550$ GeV. The measurements, which are consistent with a power law dependence of  the exclusive \jpsi photoproduction cross section, are compared to previous results from HERA and the LHC and to several theoretical models. They are found to be compatible with  previous  measurements.
\end{abstract}
\end{titlepage}

\setcounter{page}{2}

\section{Introduction}

One of the most interesting aspects of perturbative Quantum Chromodynamics (pQCD) is the evolution of the structure of hadrons in terms of quarks and gluons towards the high-energy limit~\cite{Gelis:2010nm}. Very precise measurements at HERA \cite{Aaron:2009aa,Abramowicz:2015mha} 
have shown that  the gluon structure function in protons rises rapidly at small values of the Bjorken $x$ variable,  which corresponds to the high-energy limit of QCD. This rise, interpreted as the growth of the probability density function for gluons, has to be damped at some high energy to satisfy  unitarity constraints
\cite{Gribov:1984tu}.
Although gluon saturation~\cite{Mueller:1989st} is the most straightforward mechanism to slow down the growth of the cross section, no compelling evidence for this effect has been found so far. 
Gluon saturation 
would have important implications in small-$x$ physics  and in the early stages of ultra-relativistic heavy-ion collisions at RHIC and LHC; consequently, finding evidence for gluon saturation has become a central task for present experiments and for future projects~\cite{AbelleiraFernandez:2012cc,Accardi:2012qut}.

The exclusive photoproduction of charmonium off protons (${\gamma}{\mathrm p}\rightarrow \jpsi p$) is a very clean probe with which to search for saturation effects,
because the cross section for this process depends, at leading order in  pQCD, on
the square of the gluon density in the target~\cite{Ryskin:1992ui}. The
mass of the charm quark provides a scale large
enough to allow calculations within  pQCD.
A reduction in the growth rate of the cross section for this process as the centre-of-mass energy of the photon-target system increases would signal the onset of gluon saturation effects.

This process has been extensively studied in ep interactions at the  HERA collider. Both ZEUS and H1 measured the energy dependence of exclusive \jpsi\ photoproduction off protons at $\gamma {\rm p}$ centre-of-mass energies, \Wgp,  from 20 to 305 GeV~\cite{Chekanov:2002xi,Aktas:2005xu,Alexa:2013xxa}. H1 has also measured exclusive \psip photoproduction off protons  with \Wgp between 40 and 150 GeV \cite{Adloff:2002re}.

The exclusive photoproduction of charmonium can also be studied at hadron colliders, where one of the incoming hadrons is the source of the photons and the other
the target. CDF  measured  exclusive \jpsi production at mid-rapidities in p$\bar{\rm p}$ collisions at the Tevatron \cite{Aaltonen:2009kg}, while LHCb  reported  measurements of exclusive \jpsi and \psip production at forward rapidities in pp collisions at the LHC~\cite{Aaij:2014iea}.
In addition, $\Upsilon$ production has  been studied, both at HERA~\cite{Adloff:2000vm,Chekanov:2009zz,Abramowicz:2011fa} and in pp collisions at the LHC~\cite{Aaij:2015kea}. In all cases, the experimental signature is the production of a vector meson in a collision in which there is no other hadronic activity, apart from the possible emission of a few very forward neutrons.

In hadron colliders,  as the incoming particles can each be both source or target, there are two possible centre-of-mass energies for the photon-target system. This ambiguity can be resolved
if the contribution of one beam acting as the source is much stronger than the other. This is the case 
for p--Pb collisions, because  photon emission grows with the square of the electric charge and thus emission by the Pb ion is strongly enhanced with respect to that from the proton. 
ALICE published  the first measurement of exclusive photoproduction of \jpsi, at forward and backward rapidities, in p--Pb collisions at the LHC~\cite{TheALICE:2014dwa}.

In this Letter we present the measurements of exclusive \jpsi photoproduction in collisions of protons with Pb nuclei at a\ centre-of-mass energy per nucleon pair $\sqrt{s_{\rm NN}} = 5.02$ TeV. The measurements are performed in the semi-backward ($-2.5<y<-1.2$), mid ($|y|<0.8$) and semi-forward ($1.2<y<2.7$) rapidity intervals, where the rapidity $y$ of the \jpsi is measured in the laboratory frame with respect to the direction of the proton beam.
The value of \Wgp is determined by the rapidity of the \jpsi according to  $W_{\gamma {\rm p}}^2 = 2 E_\mathrm{p} M_{\jpsi} \exp (-y)$, where $M_{\jpsi}$ is the \jpsi mass and $E_\mathrm{p} = 4$ TeV is the energy of the proton beam.
Thus, the measurements presented here span  the \Wgp range from 40 to 550 GeV. Together with the measurements presented in~\cite{TheALICE:2014dwa} (for the rapidity ranges $-3.6 < y < -2.6$ and $2.5 < y < 4$),  ALICE data map the \Wgp dependence of the cross section for this process from about 20 to 700 GeV, which, using the standard definition $x=M^2_{\jpsi}/W^2_{\gamma {\rm p}}$, corresponds to three orders of magnitude, from  ${\sim}2\times 10^{-2}$ to ${\sim}2\times 10^{-5}$.

\section{Experimental set-up}

The ALICE detector and its performance are described in detail in~\cite{Aamodt:2008zz,Abelev:2014ffa}. It consists of two main sections: a central barrel placed in a large solenoid magnet ($\mathrm{B} = 0.5$\ $\mathrm{T}$), covering the pseudorapidity region $|\eta|<0.9$, and a muon spectrometer 
covering the range
$-4.0 < \eta < -2.5$. In addition, the analyses presented here make use of two other detector systems, V0 and ZDC, which are placed near the beam pipe and cover large pseudorapidities.

Because the proton and Pb beams had different energies, 4 TeV and 1.58 TeV per nucleon respectively,
the nucleon-nucleon centre-of-mass  was shifted, with respect to the laboratory frame,  by $\Delta y_{\rm{NN}} = 0.465$ in the direction of the proton beam.
Collisions were performed in two configurations, denoted by \pPb and \Pbp below, by reversing the directions of the LHC beams. According to the naming convention the {\em first}  named particle (e.g. `p' in \pPb) is the one where the beam travels from the interaction region towards the muon spectrometer. Note that the convention in ALICE is that the pseudorapidity is negative in the direction of the muon spectrometer. In contrast, the laboratory rapidity is measured with respect to the direction of the proton beam and changes sign accordingly for the \pPb and \Pbp  configurations. In the semi-backward and semi-forward analyses, the negative rapidity points come from the \Pbp data, and the positive rapidity points from the \pPb.

The ALICE apparatus consisted of 18 detector systems at the time the data used in this study were taken. The detectors relevant for the analysis are described in more detail below.

\subsection{The ALICE central barrel}

The innermost component of the central barrel is the Inner Tracking System (ITS)~\cite{Aamodt:2010aa}. The ITS contains six cylindrical layers of silicon detectors, with the innermost layer at a radius of 3.9 cm with respect to the beam axis and the outermost layer at 43 cm. The two rings closest to the beam form the  Silicon Pixel Detector (SPD) and cover  the  pseudorapidity ranges
$|\eta| < 2$ and $|\eta| < 1.4$, for the inner  and outer  layers respectively. It is a fine granularity detector, having about $10^{7}$ pixels. In addition to functioning as a tracking device, the SPD signals can be used to build triggers. The other four layers, two employing silicon drift chambers and two silicon microstrips, are used in this analysis exclusively for tracking.

The Time Projection Chamber (TPC), surrounding the ITS, is a large cylindrical detector used for tracking and for particle identification~\cite{Alme:2010ke}. Its active volume  extends from 85 to 247 cm in the radial direction and  has a total length of 500 cm longitudinally, centred on the interaction point. A 100 kV central electrode separates its
 two drift volumes, providing an electric field for electron drift. The two end-plates are instrumented with Multi-Wire-Proportional-Chambers (MWPCs) with 560 000 readout pads in all, allowing high precision track measurements in the transverse plane. The $z$~coordinate is given by the drift time in the TPC electric field.  The TPC acceptance covers the pseudorapidity range $|\eta| < 0.9$ for tracks fully traversing it.
Ionization measurements made along track clusters are used for particle identification.

Beyond the TPC lies the Time-of-Flight detector (TOF)~\cite{Akindinov:2013tea}. It is a large cylindrical barrel of Multigap Resistive Plate Chambers (MRPCs) giving very high precision timing for tracks traversing it. Its pseudorapidity coverage matches that of the TPC. It is also capable of delivering signals for triggering purposes~\cite{AKINDINOV2009372}.

\subsection{The ALICE muon spectrometer}

The muon spectrometer consists of a 3 Tm dipole magnet, coupled with a tracking and a triggering system. A ten interaction-length conical front absorber, made of carbon, concrete and steel, is placed between the interaction point (IP) and the
muon spectrometer tracking system, between 0.9 and 5 m from the IP, to filter out primary hadrons. Muon tracking is performed by means of five tracking stations, each one made of two planes of Cathode Pad Chambers. A 7.2 interaction-length iron wall is placed after the tracking stations. It is followed by the muon trigger system, based on two stations equipped with Resistive Plate Chambers. The trigger system can provide single-muon and dimuon triggers with a programmable \pt threshold. The \pt threshold is an approximate value owing to the coarse grain of the trigger spatial information. In addition, there is a conical absorber made of tungsten, lead and steel, that surrounds  the beam pipe at small angles ($\theta < 2^{\circ}$) and shields the spectrometer from secondary particles produced in interactions of primary particles in the beam pipe.

\subsection{The V0 and the ZDC systems}

Two sets of hadronic Zero Degree Calorimeters (ZDC) are located at 112.5 m on either side
of the interaction point. Each station consists of two detectors, one to detect protons and the other neutrons. The neutron ZDCs detect neutrons emitted in the very forward region ($|\eta|>8.7$). The
calorimeters detect the Cherenkov light produced in quartz fibers by the hadronic showers. Each detector is read out by five photomultipliers. Their timing  and energy resolution allow one to select events with neutrons produced at very forward rapidity, as could be the case in events with proton dissociation.

 The V0 counters~\cite{Abbas:2013taa} consist of
two arrays of 32 scintillator cells each, covering the range 2.8~$<$ $\eta$ $<$~5.1
(V0A on the side opposite to the muon spectrometer) and --3.7~$<$~$\eta$~$<$~--1.7
(V0C, on the same side as the muon spectrometer) and positioned respectively at $z$~= 340~cm and $z$~= --90~cm
from the interaction point. The raw signals are used in the trigger, but are refined offline for use in the final data selections. For example, these detectors provide\ timing information with a resolution better than 1 ns, which\ can be used to distinguish two different timing windows, one corresponding to the passage of particles coming from collisions, and the other to particles coming from upstream interactions with residual gas molecules, termed ``beam-gas'' interactions.\

\subsection{Triggers}
\label{section:trigger}

Two different signatures have been used in the analyses described below. In one case, the  mid-rapidity analysis, the \jpsi candidate decays into a pair of leptons (either $e^+e^-$ or $\mu^+\mu^-$), both of which are detected with  the central barrel detectors.
The second case, the semi-forward (or semi-backward) analysis, corresponds to a \jpsi candidate decaying into a pair of muons where one of them is measured with the muon spectrometer and the other with the central barrel detectors.

The trigger for the mid-rapidity analysis required at least two, and at most six, fired pad-OR in the TOF detector~\cite{Akindinov:2013tea}, such that at least two of the signals came from modules separated by more than 150 degrees in azimuth,
{since the \jpsi decay products are predominantly produced back-to-back in azimuth. The trigger also required no signal  in V0A and no signal in V0C, to ensure that there is very little extra  hadronic activity in the event, and at most six fired pixel chips in the outer layer of the SPD. In addition there were required to be two inner-outer pairs of these pixels having a back-to-back topology.

 Three   triggers were used for the semi-backward and semi-forward analyses. The first was active during the \pPb data taking period. It required a muon candidate in the muon spectrometer trigger system with a transverse momentum above 0.5 GeV/{$c$\ (as assessed in the trigger system electronics), at most 4 cells with signal in the V0C and\ no activity in the V0A. The SPD requirement was to have at least 1 pixel hit in either the\ inner or the\ outer layer and less than 7\ pixels fired in the outer layer.
Two different triggers were active during the \Pbp period. For the first one, in addition to the previous conditions,\ it was also required that at least one cell of the V0C had a signal. The second one had the same requirements as the first, but in addition vetoed signals from V0A compatible with beam-gas timing. These tighter conditions were required owing to an increased background trigger rate.

The integrated luminosity {$\mathcal{L}_\mathrm{int}$} was corrected for the probability that the exclusivity requirements could be spoilt by multiple interactions in the same bunch crossing. This pile-up correction is on average 5\%. The total luminosity of the sample used for the mid-rapidity analysis is 2.1 nb$^{-1}$ (4.8 nb$^{-1}$) in the \pPb (\Pbp) period, while for the semi-forward (semi-backward) analysis  it is 3.1 nb$^{-1}$ (3.7 nb$^{-1}$) in the  \pPb (\Pbp) period.\ Note that in the mid-rapidity case, the \pPb and \Pbp data samples have been analysed together, taking into account the inversion of the rapidity sign, 
to increase the size of the central rapidity data sample.

\section{Data samples}
\subsection{Event and track selection for the mid-rapidity analysis}

The events used in the mid-rapidity analysis fulfilled the following criteria.
\begin{enumerate}
\item They fired the corresponding  trigger.
\item There were exactly two tracks reconstructed with the TPC and the ITS.
\item The primary vertex had at least two tracks defining it.
\item Each track had at least 50  TPC space points out of a maximum of 159, a TPC-$\chi^2/\mathrm{dof}<4$, at least one SPD point and pseudorapidity $|\eta|<0.9$.
\item The distance of closest approach of each track to the (nominal) primary vertex (DCA) was less than 2 cm in the longitudinal direction and less than $0.0182 + 0.035/\pt^{1.01}$ cm, with \pt in GeV/$c$, in the transverse direction\cite{Abelev:2014ffa}.
\item Neither track was associated with a kink candidate.
\item The coordinate of the primary vertex of the interaction along the beam-line direction was within 15 cm of the nominal interaction point.
\item The two tracks had  opposite electric charge.
\end{enumerate}
 To build dilepton candidates the tracks were separated into electrons and muons using the energy loss measured by the TPC, such that:
  \begin{enumerate}
 \item the quadratic sum of the normalised deviation from the corresponding Bethe-Bloch expectation for the positive and negative track had to be less than 16;
 \item the rapidity, measured in the laboratory frame, of the dilepton candidate, formed by the 4-vector sum of the two tracks with the corresponding mass hypothesis, had to be in the range $|y|<0.8$.
 \end{enumerate}

In addition, the events had to

\begin{enumerate}
\setcounter{enumi}{2}
\item have no signal in V0  as determined by\ the offline decision, and
\item no signal in either of the neutron\ ZDCs.
\end{enumerate}

\subsection{Event and track selection for the semi-backward and semi-forward analyses}
The events used in the semi-backward and semi-forward analyses fulfilled the following criteria.
\begin{enumerate}
\item They fired the corresponding  trigger.
\item There was exactly one track reconstructed with the TPC and the ITS.
\item This track fulfilled similar requirements to those for the central analysis tracks, with two modifications: it had to be compatible with a muon (within $4\sigma$)\ according to the energy-loss measurement, and the acceptance was enlarged to $|\eta|<1.1$.
\item Furthermore, the distance of closest approach of this track to the nominal interaction point had to be less than 15 cm in the direction along the beam-line.
\item The event also had to have exactly one\ track in the muon spectrometer such that:
\begin{enumerate}
\item the track is matched to a trigger track above the \pt threshold;
\item the radial coordinate of the track at the end of the front absorber ($R_{\rm abs}$) was required to be in the range  17.5 cm $< R_{\rm abs} <$ 89.5 cm;
\item it had a pseudorapidity in the range $-4.0 < \eta < -2.5$ ($-3.7 < \eta < -2.5$) in the \pPb (\Pbp) period  and fulfilled the so called $p\times$DCA requirement.

The $p\times$DCA requirement uses the difference in the distributions for signal and beam-induced background in the ($p$, DCA) plane, where $p$ is   the track momentum and DCA is the distance between the interaction vertex and the track extrapolated to the vertex transverse plane. If in a given event no interaction vertex was found, the point (0, 0, 0) was used to calculate the distance of closest approach.\ The pseudorapidity\ ranges are different in \pPb and \Pbp owing to the additional V0C trigger requirement (see section \ref{section:trigger}).
\end{enumerate}

\item The two tracks had to have opposite electric charge.
\item The rapidity, measured in the laboratory frame, of the dimuon candidate, formed by the 4-vector sum of the two tracks with the muon-mass hypothesis, had to be in the range $1.2 <y<2.7$ for the \pPb and $-2.5<y<-1.2$ for the \Pbp sample, respectively.
\item In addition, the events had to have no signal in V0A, at most one cell with a signal in V0C,
(in both cases as determined by the offline reconstruction),
 no signal in either of the neutron\ ZDCs and at most one tracklet in the SPD layers.
 \end{enumerate}

\subsection{Selection of background samples}
\label{Sec:BkgdSample}
Several background samples, one for each analysis, were selected. 
The events in these samples satisfy the conditions listed in the previous sections, and in addition there is required to be a signal, compatible with the presence of a neutron, in the ZDC placed in the proton direction. Both like-sign and opposite-sign dilepton candidates are accepted.

These samples are highly efficient to tag dissociative processes (where the target proton breaks up) and other hadronic collisions and were used to build templates to subtract the remaining background in the real data samples as explained below.\ As the trigger suppresses events with higher \pt dileptons ($\pt \ge 0.5$ GeV/$c$), due to the condition on azimuthal separation, the number of events\ in this \pt range is limited. Therefore, each background spectrum\ was used to generate a high statistics sample, which was smoothed to produce the final template. This is referred to as ``non-exclusive background'' in the analysis of the \pt spectrum. The selections and smoothing algorithms were varied to estimate the systematic uncertainties, and the background \pt\ distribution was verified independently by selecting hadronic activity in V0C instead of the ZDC.

\subsection{Monte Carlo samples}
\label{Sec:MCSample}

The STARLIGHT Monte Carlo generator~\cite{Ref:SLMC,Klein:2016yzr} was used to generate large samples for the following processes:  exclusive $\jpsi$ photoproduction off proton ($\gamma{\mathrm{p}}\rightarrow \jpsi\mathrm{p}$) and Pb ($\gamma{\mathrm{Pb}}\rightarrow \jpsi\mathrm{Pb}$)  targets, exclusive photoproduction of $\psip$ decaying into $\jpsi+X$, and continuum dilepton production $\gamma\gamma\rightarrow l^+l^-$. The physics models used by STARLIGHT are described in~\cite{Klein:2016yzr,Klein:1999qj,Baltz:2009jk}. All generated events were passed through a full simulation of the detector based on GEANT 3 transport code~\cite{Brun:1987ma} and subjected to the same analysis chain as real data. The Monte Carlo took into account the changing status of the detectors during data-taking, and their residual misalignment.

In all cases the polarization of the photoproduced charmonium was taken as transverse, as expected for photoproduction of vector mesons. For the $\psip$ sample three cases were studied, corresponding to an assumption of longitudinal, transverse and no polarisation of the $\jpsi$ produced by the decay of the $\psip$.

The dependence of exclusive photoproduction on the momentum transferred in the target vertex is modelled in STARLIGHT using an exponential distribution $F(Q^{2})=e^{-bQ^{2}}$, with the slope $b$ taking the default value of 4 GeV$^{-2}$\cite{PhysRevLett.84.2330}. HERA measurements have shown that the value of this parameter depends on \Wgp~\cite{Aktas:2005xu}. This effect is particularly important for the mid-rapidity and the semi-backward samples, where \Wgp is larger. To take this into account  additional samples of exclusive $\jpsi$ were produced using STARLIGHT with a value of  $b$  given by the H1 parameterisation from \cite{Aktas:2005xu}.

\section{Data analysis}

\subsection{Signal extraction}
The mass distributions for the dilepton candidates that passed  the selection described in the previous section are shown in Fig. \ref{fig:MassDistribution}. In all cases a clear \jpsi signal is  observed over a small background. The data can be satisfactorily described in all cases by a combination of a Crystal-Ball function (CB)~\cite{Skwarnicki:1986xj} and an exponential distribution, which represent the signal and the background from continuum dilepton production, respectively. The tail parameters of the CB distribution have been fixed to the values obtained by fitting simulated events.

Figure \ref{fig:PtDistribution} shows the distribution of transverse momentum for dilepton pairs having mass ($m_{\mathrm{l^+l^-}}$) in the range  $2.9 < m_{\mathrm{l^+l^-}}< 3.2$ GeV/$c^2$ for all the samples, except for the dimuon case in the mid-rapidity analysis where the lower bound was raised to 3.0 GeV/$c^2$ to make use of the better mass resolution of this channel. In all cases the transverse momenta populate the region below about 1 GeV/$c$, as expected from exclusive photoproduction off protons. These distributions are used to extract the number of exclusive \jpsi, within the stipulated mass range, in each of the measured rapidity intervals. An extended binned likelihood method is used in  fits of data to a sum of templates for signal and background obtained from data (see Section~\ref{Sec:BkgdSample}) and from Monte Carlo (see Section~\ref{Sec:MCSample}) samples. These templates are also shown in the figure.

The  number of candidates from the continuum dilepton template, which is a free parameter in the fit,  was found always to be less than what is obtained, in the given mass range, from the exponential part of the fit to the mass distribution. This exponential distribution includes both the exclusive and the dissociative dilepton production, while the  template describes only the exclusive component. The  number of candidates for coherent \jpsi production off lead has been estimated using STARLIGHT. The STARLIGHT predictions were rescaled so as to be compatible, within the uncertainties, with ALICE measurements in Pb--Pb collisions~\cite{Abelev:2012ba,Abbas:2013oua}. The normalisation of the exclusive \jpsi and non-exclusive background templates have been left free.\ Note that the non-exclusive background is lower at higher energies of the photon-target system, as has already been observed at HERA \cite{Aktas:2005xu,Alexa:2013xxa}, and also by ALICE \cite{TheALICE:2014dwa}.

The number of candidates obtained from the fitted template for exclusive \jpsi production has been corrected for the feed-down from exclusive production of \psip decaying into $\jpsi+X$ \cite{Abbas:2013oua}.  This  correction amounts to about 2\% (4\%) for the semi-backward and semi-forward (mid-rapidity) analyses.
The correction factor to take into account the acceptance and efficiency ($\eXa$) of the detector was computed using  simulated events,
as described in section \ref{Sec:MCSample}. This factor depends on the rapidity of the \jpsi and varies from around 3\% to almost 9\%. For the 
semi-backward sample the $\eXa$ includes the efficiency of the V0C (around 60\%) to detect the passing muon (because it is included in the trigger).
The values for the number of exclusive \jpsi extracted in the different data samples, as well as the corresponding \eXa factors and available luminosity, are summarised in Table~\ref{Tab:Inputs}.\ Note that, in order to make best use of the available statistics, both the central and the semi-forward (but not the semi-backward) samples have been sub-divided into two rapidity bins.

 \begin{table}

\begin{center}
\begin{tabular}{cccc}
\hline
 Rapidity  & $N_{\jpsi}$ & \eXa & $\mathcal{L}_{\mathrm{int}}$ (nb$^{-1}$) \\ \hline
(1.9,2.7)   &  73$\pm$15 & 0.071 & 3.147    \\
(1.2,1.9) & 98$\pm$17 & 0.086 & 3.147  \\
(0.0,0.8)  & & & \\
 dielectron &   106$\pm$16 & 0.029 & 6.915 \\
  dimuon &   196$\pm$17  & 0.057 & 6.915 \\
(-0.8,0.0)  & & & \\
 dielectron  &    117$\pm$13 & 0.033 & 6.915  \\
  dimuon &    199$\pm$20  & 0.063 & 6.915 \\
(-2.5,-1.2) &   76$\pm$11 & 0.037 & 3.743 \\
\hline
\end{tabular}
\end{center}
\caption{Number of measured exclusive \jpsi, value of the efficiency and acceptance correction factor, as well as integrated luminosity for each rapidity interval. These values are used in  Eq.~\ref{Eq:XS} to compute the measured cross sections.
}
\label{Tab:Inputs}
\end{table}

\begin{figure}[t]
\begin{center}$
\begin{array}{cc}
\includegraphics[width=0.47\textwidth]{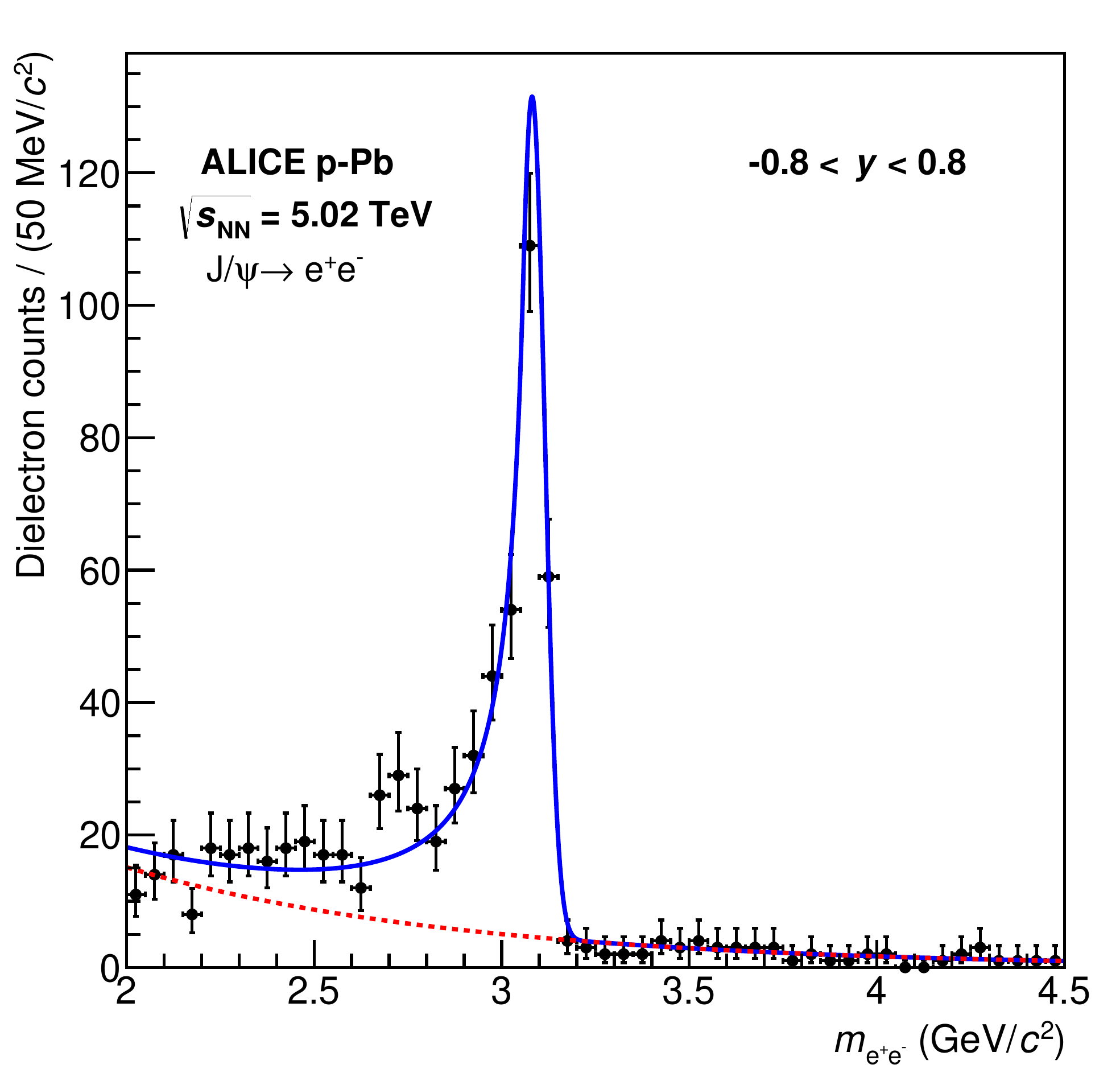} &
\includegraphics[width=0.47\textwidth]{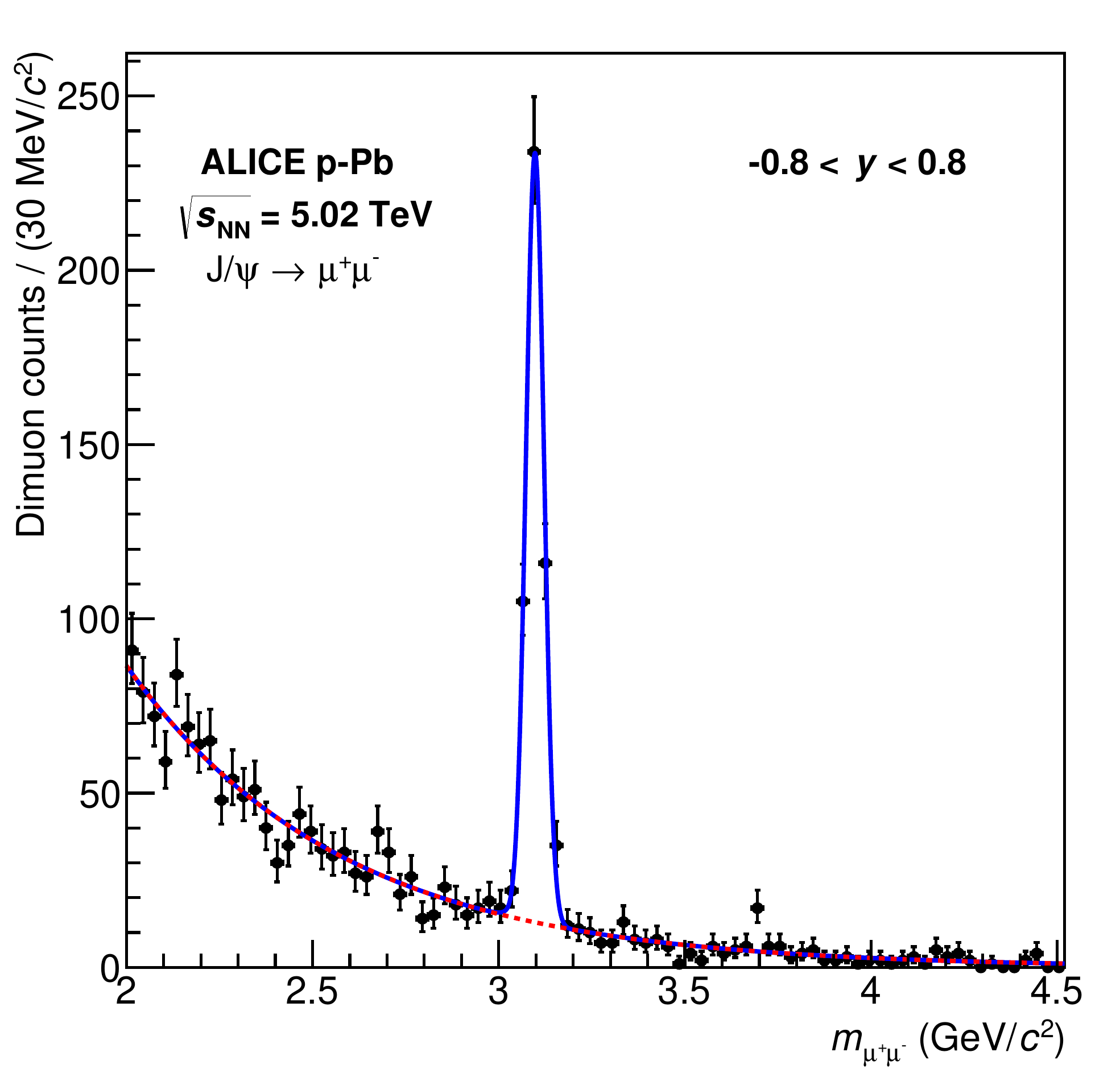}
\\
\includegraphics[width=0.47\textwidth]{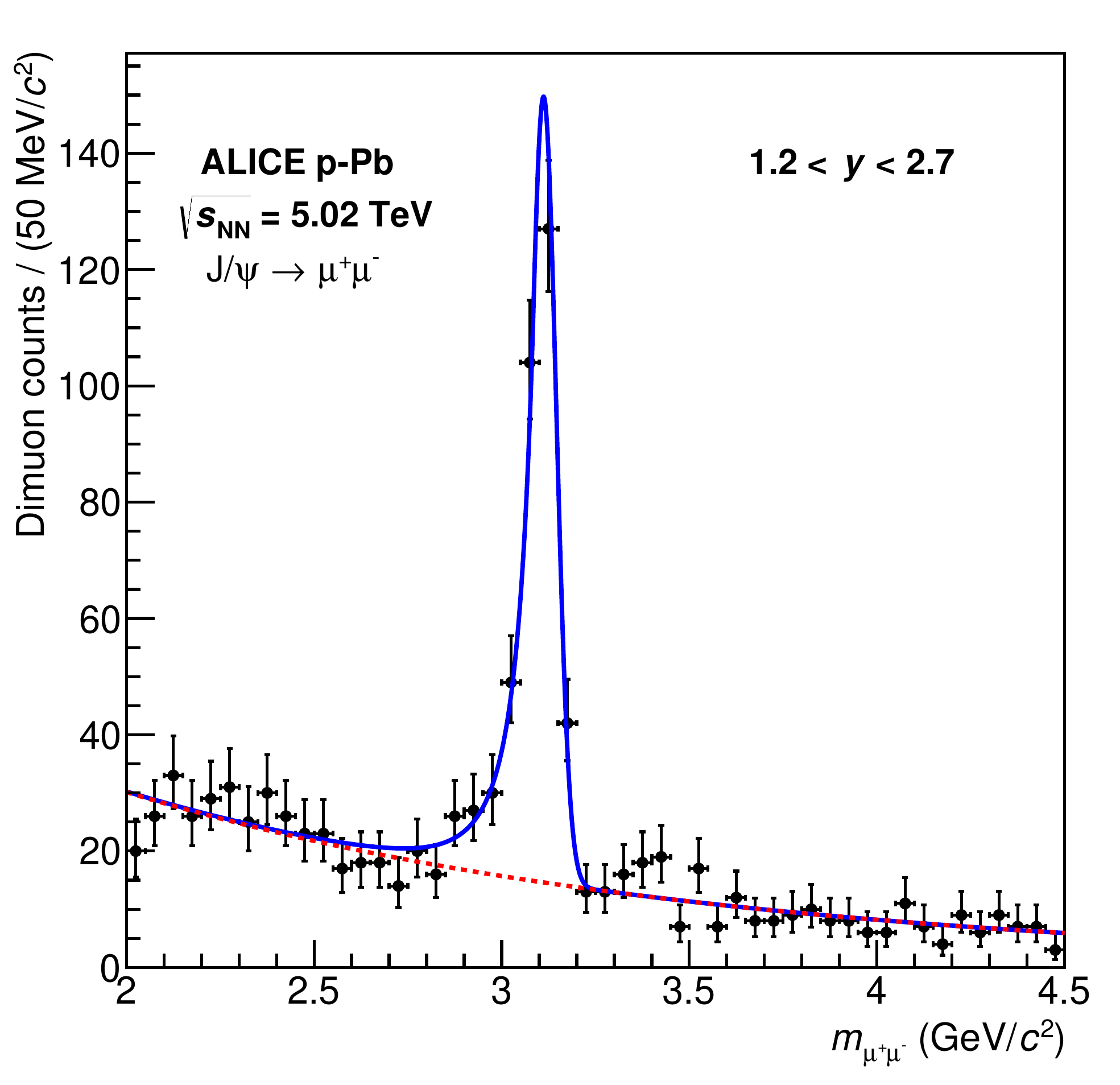} &
\includegraphics[width=0.47\textwidth]{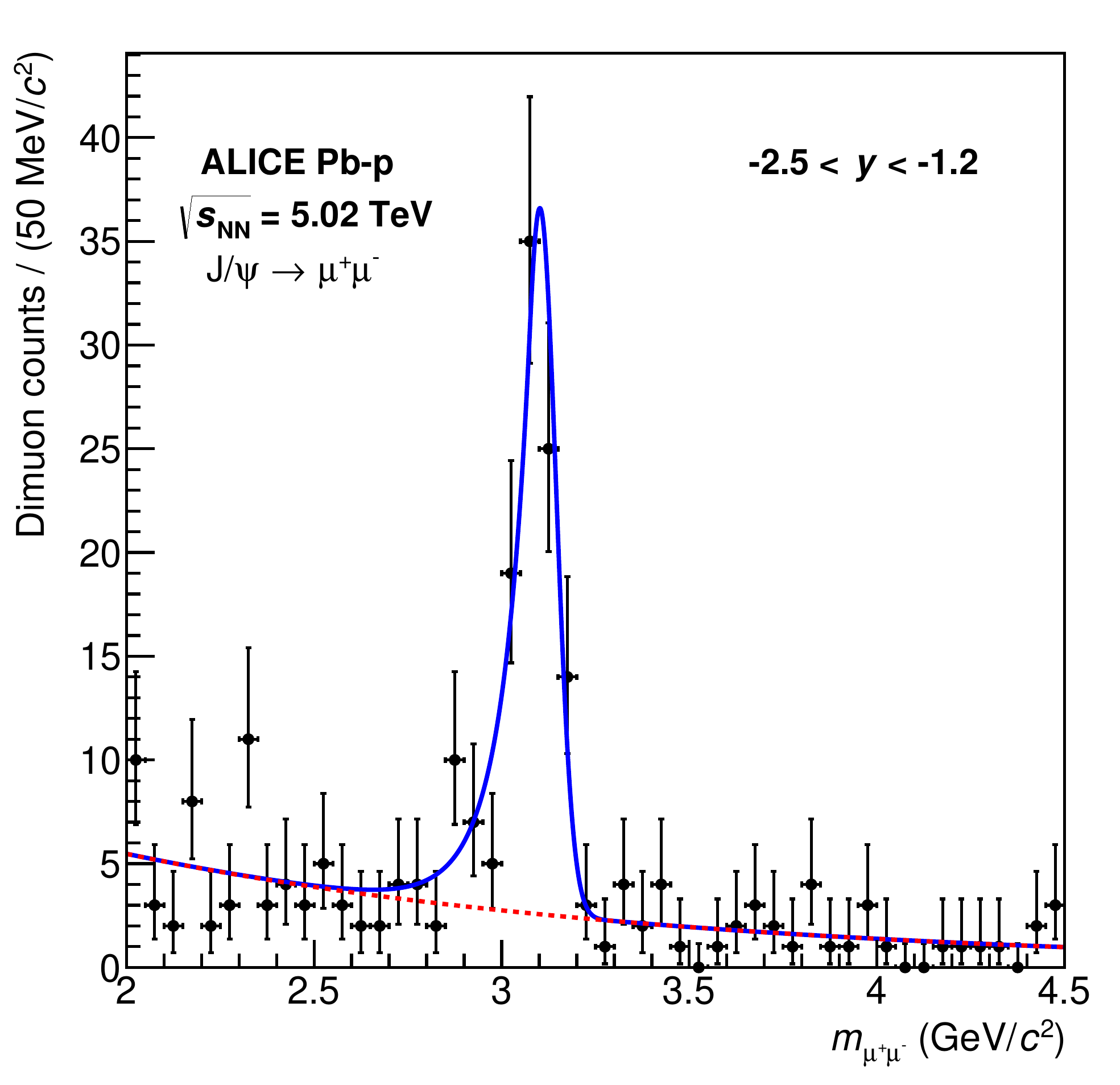}
\end{array} $
\end{center}
\caption{Mass distributions of selected dileptons for the dielectron (upper left) and dimuon (upper right) samples for the central analysis and dimuon samples for the semi-forward   (lower left) and semi-backward (lower right) analyses. In all cases the data are represented by points with error bars.
The solid blue line is a fit to a Crystal-Ball function plus an exponential distribution, where this last contribution is shown by a dotted red line.}
\label{fig:MassDistribution}
\end{figure}	

\begin{figure}[t]
\begin{center}$
\begin{array}{cc}
\includegraphics[width=0.48\textwidth]{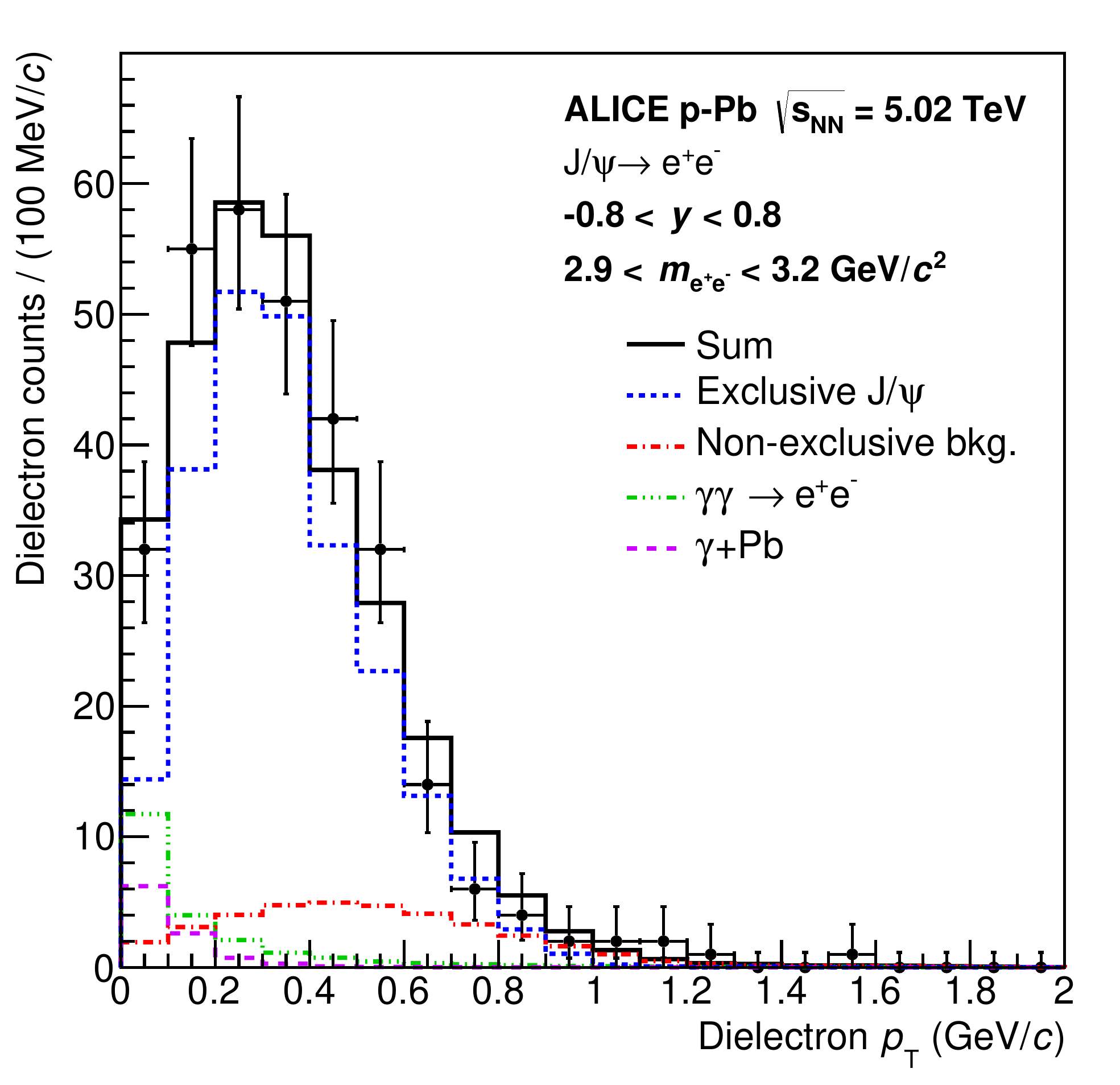} &
\includegraphics[width=0.48\textwidth]{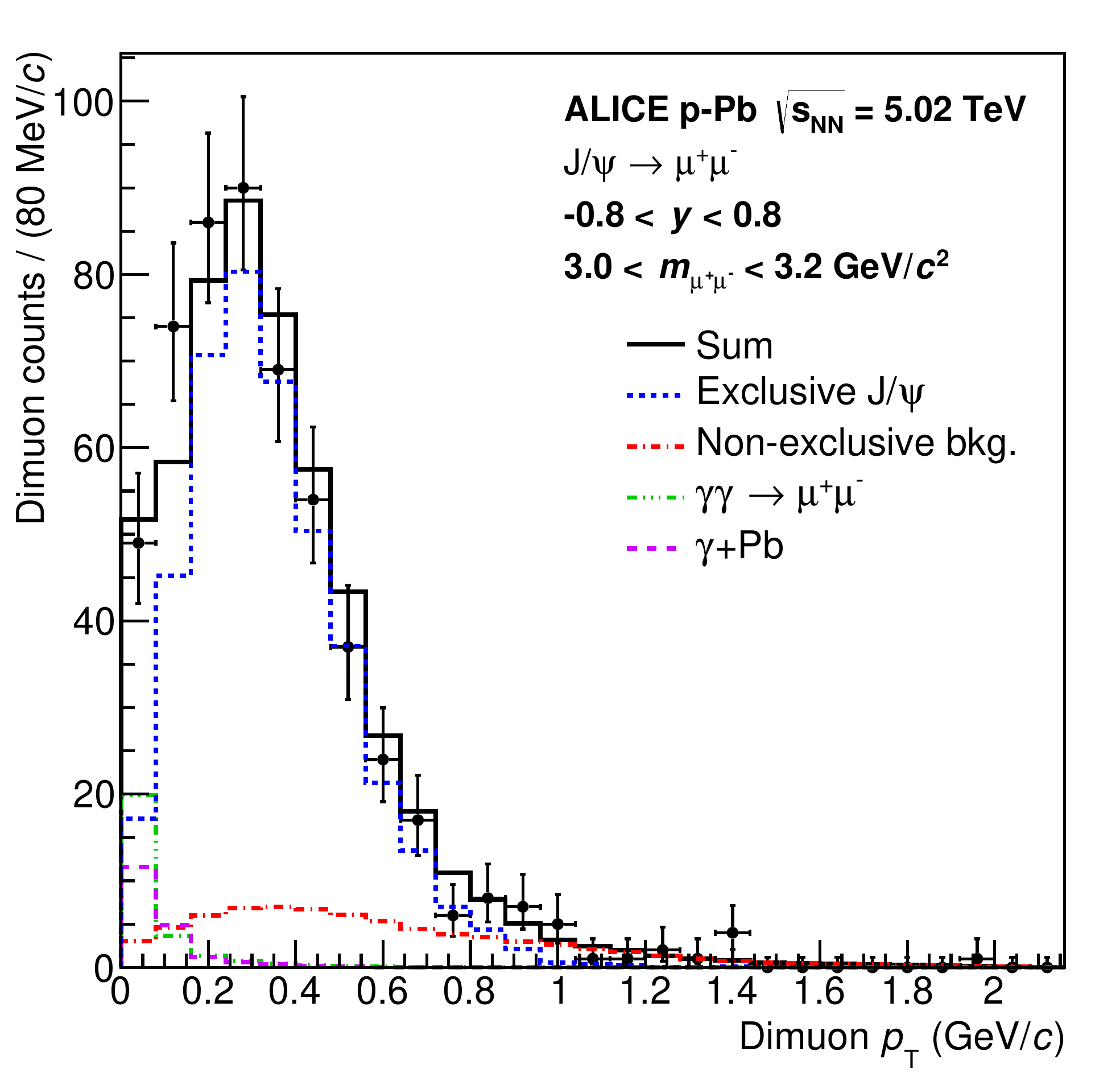}
\\
\includegraphics[width=0.48\textwidth]{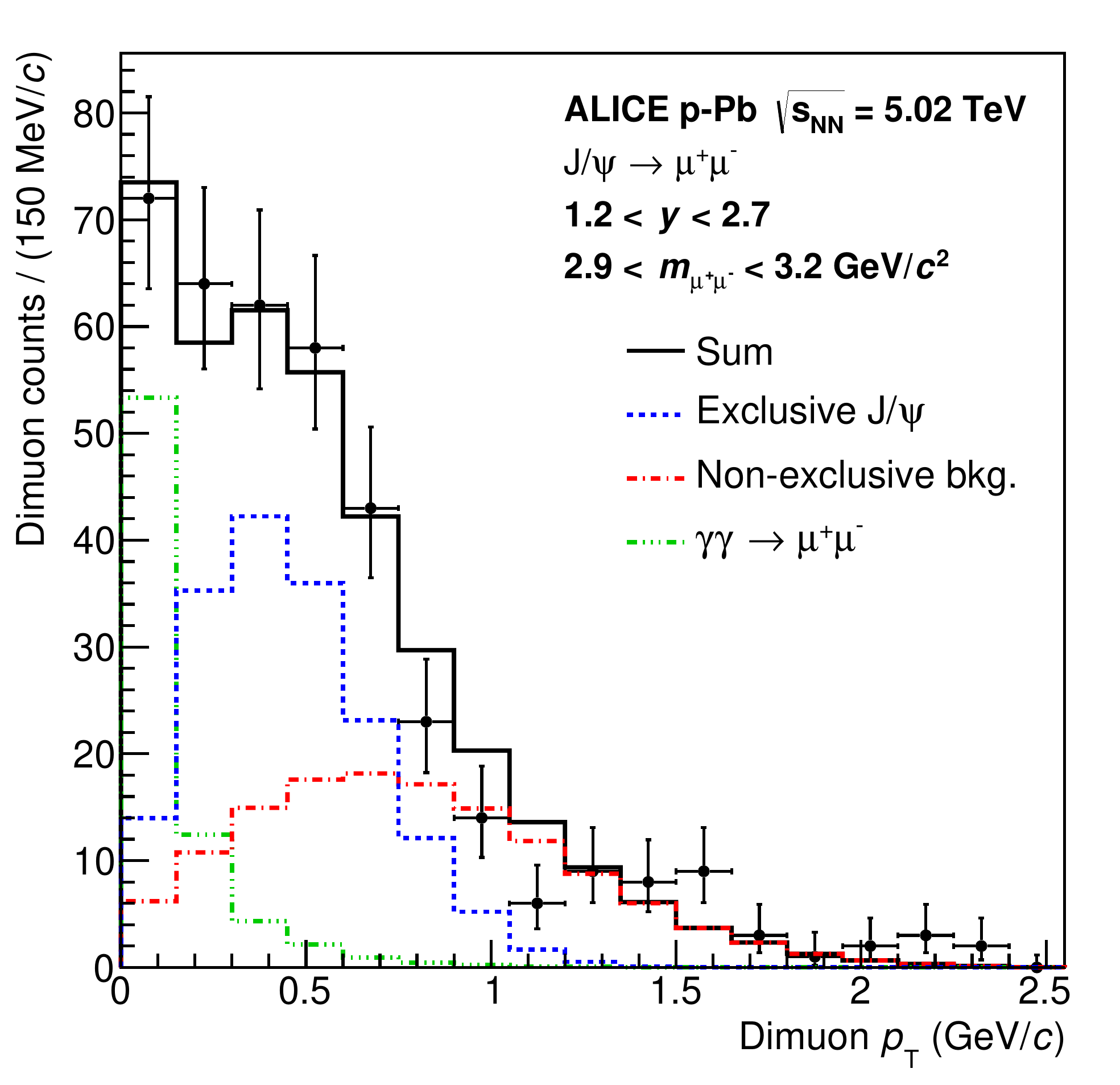} &
\includegraphics[width=0.48\textwidth]{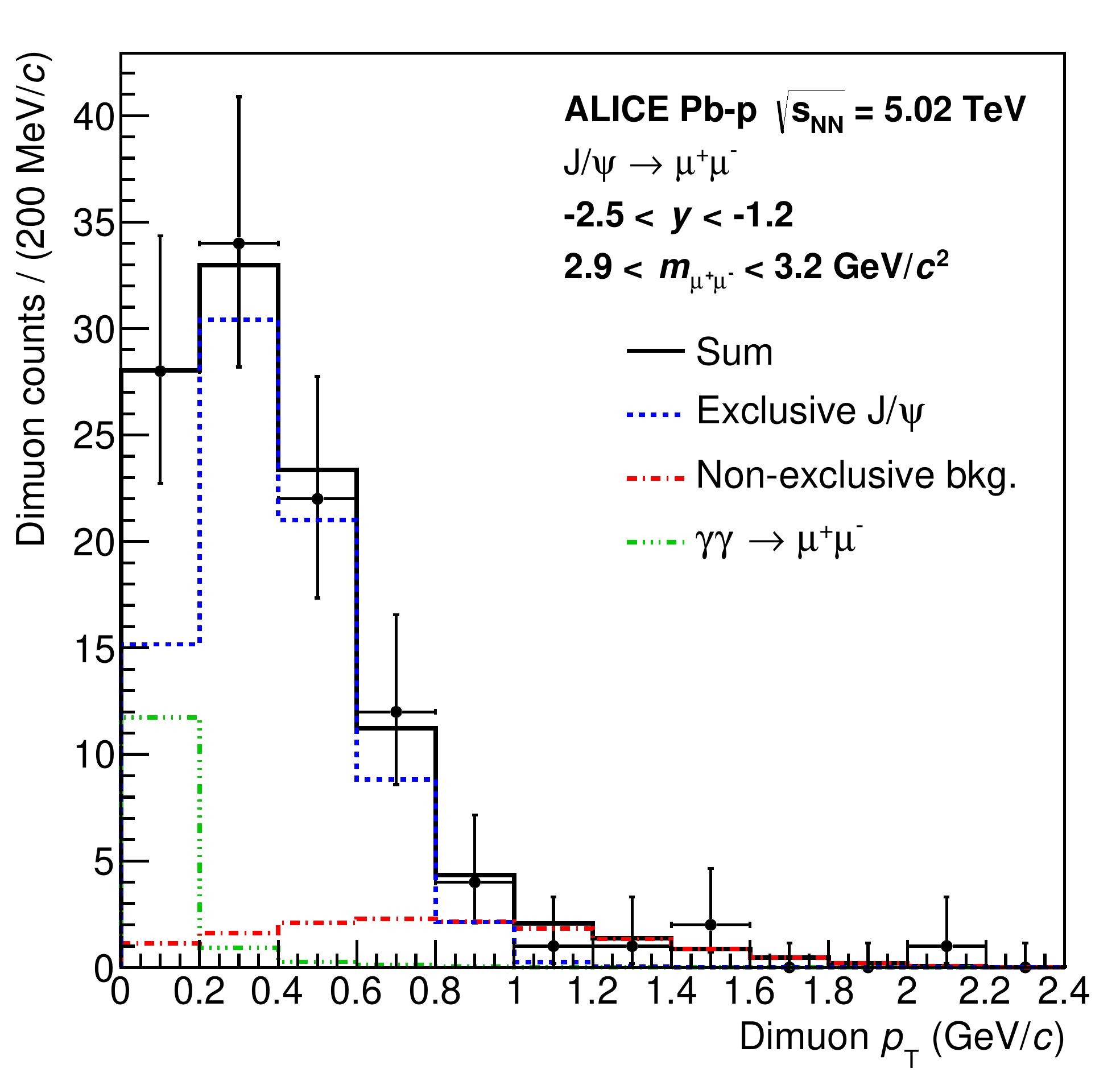}
\end{array} $
\end{center}
\caption{ Transverse momentum distributions of dileptons around the \jpsi mass  for the dielectron (upper left) and dimuon (upper right) samples for the central analysis and dimuon samples for the semi-forward   (lower left) and semi-backward (lower right) analyses. In all cases the data are represented by points with error bars. The  blue, magenta (dash) and green (dash-dot-dot) lines correspond to Monte Carlo templates for \jpsi\ coming from exclusive photoproduction off protons or off lead and continuous dilepton production respectively. The red (dash-dot) line is a template for dissociative and hadronic background obtained from data. The solid black line is the sum of all contributions. }
\label{fig:PtDistribution}
\end{figure}	

\subsection{Estimation of systematic uncertainties}

Several sources of systematic uncertainty have been studied. They are discussed below, while their contribution to the uncertainty on the measured  cross sections are summarised in Table~\ref{Tab:SU}.

 The uncertainty on the tracking efficiency in the TPC was estimated repeating the analyses using different track selections. Four different values for the minimal number of points and three values for the minimum number of TPC pads crossed by the track were used. The systematic uncertainty related to the  selection of  tracks at mid-rapidity was estimated from the spread of the variations with respect to the standard selection. It varies from  0.8\% to 5.7\%.

The uncertainty related to the identification of electrons and muons using their energy deposition in TPC has been obtained using an alternative selection based directly on the energy loss\ \cite{Abelev:2013yxa}. The difference between the two methods was used as an estimate of the systematic uncertainty, and was at most 1.3\%. Cross contamination from muon pairs in the electron sample and vice versa was found to be negligible.

The uncertainty on the single muon tracking efficiency in the muon spectrometer was obtained by comparing the results of measurements performed on simulations with those from real data~\cite{Abbas:2013oua} and amounts to 2\% (3\%) for the \pPb (\Pbp) period. There is also a 0.5\% contribution from variations on the conditions required to match the trigger and the tracking information of a given muon.\

The uncertainties related to triggering in the muon spectrometer have been evaluated  as
in~\cite{Abelev:2013yxa}. The efficiency maps\ of the trigger chambers have\ been obtained using data. The statistical uncertainty on this procedure has\ been used to vary the efficiency in simulations, which was then used to estimate a systematic uncertainty of 1\%. There is also a small discrepancy between the efficiency in data and in simulations around the trigger threshold.  This gives a contribution of 1.7\% (1.3\%) for the \pPb (\Pbp) period.
The addition in quadrature of these two effects yields the uncertainty on muon triggering.

The two main contributions to the uncertainty on the trigger efficiency for the mid-rapidity analysis come from the back-to-back topology condition in the SPD and TOF detectors. The first one was estimated using a data sample obtained using a zero-bias trigger (all bunch crossings taken). It amounts to 4.5\%. The second one was taken from the analysis of \cite{Abbas:2013oua}. It amounts to  (--9\%, +3.8)\%,
using a zero bias\ trigger to compute the efficiency  and comparing the result with the efficiency from simulated events. 

The efficiency of the V0C to detect one muon from the \jpsi decay, which is needed in the semi-backward analysis, was estimated using events from the \pPb period, whose trigger did not include this V0C requirement. The procedure was cross checked using the forward dimuon sample used for the analysis described in~\cite{TheALICE:2014dwa}. The efficiency depends slightly on the mass range used to compute it. The addition in quadrature of the statistical uncertainty (2.7\%) and the mass dependence (2.0\%) yields an uncertainty of 3.4\%. The veto efficiency of the V0 detector can also be estimated offline using a more complex algorithm than that used in the online trigger. To estimate the uncertainty on the use of V0 to veto extra activity, the analyses were\ compared using the online V0 decision only and requiring (standard selection) in addition the offline decision, with the difference giving the systematic uncertainty.\  The uncertainty varies from 1.2\% to 3.5\%.

The trigger conditions associated with the upper limits in the activity in TOF, SPD and V0C have a negligible effect on the systematic uncertainty, because the limits are set well above the levels of activity produced by the signal in our sample.

The systematic uncertainty on the yield was obtained by varying the range of fit to the transverse momentum template, the width of the binning and the selections and smoothing algorithms used to determine the non-exclusive template. (See section \ref{Sec:BkgdSample}.) Furthermore, the value of the $b$ parameter used in the production of the exclusive \jpsi template was varied, taking into account the uncertainties\ reported by H1~\cite{Aktas:2005xu}. The uncertainty varies from 1.9\% to 3.6\%. (See ``signal extraction'' in Table \ref{Tab:SU}.)

The polarization of the \jpsi coming from \psip feed-down is not known. The uncertainty on the amount of feed-down has been estimated by assuming that the \jpsi was either not polarised or that it was fully transversely or fully longitudinally polarised. This uncertainty is asymmetric and varies from $+1.0\%$\ to $-1.4\%$. (See ``feed-down'' in Table \ref{Tab:SU}.)

The uncertainty on the measurement of the luminosity has a  contribution of 1.6\%, which is correlated between the \pPb and \Pbp data-taking periods and, in addition, an uncorrelated part of 3.3\% (3.0\%) in the \pPb (\Pbp) configuration~\cite{Abelev:2014epa}. For the mid-rapidity analysis, which has data from both p--Pb and Pb--p, the uncorrelated part of the uncertainty on the luminosity amounts to 2.3\%.

Furthermore, the uncertainties on the TPC selection, particle identification, offline V0 veto efficiency and signal extraction  are uncorrelated across rapidity. The uncertainty on the \psip feed-down is correlated for all rapidities.  The uncertainties on muon tracking, matching and trigger efficiencies are mostly uncorrelated across rapidity, and\ for the purposes of this analysis we treat the uncertainties as fully uncorrelated.\ (This includes also the measurements from~\cite{TheALICE:2014dwa}.) The trigger efficiency at mid-rapidity is correlated between the (0.0,0.8) and (-0.8,0.0) rapidity intervals.

The mid-rapidity analysis offers two other possibilities to cross check the consistency of the results. One can compare the results of the cross section in the \pPb and \Pbp periods, and one can compare the results of the electron and the muon decay channels. The cross sections agreed in all cases within the statistical uncertainties.\  As a cross-check, the possible contribution from incoherent production of \jpsi off the Pb nucleus was investigated, and found to be negligible. In addition, the effect on the  \eXa correction\ of varying the slope parameter {\it b} within the HERA limits was also found to be negligible.

Systematic effects related to noise or pileup events in the ZDC are estimated using randomly triggered events and are also found to be negligible.

\begin{table}

\begin{center}
\begin{tabular}{lccccc}
\hline
     & (1.9,2.7) & (1.2,1.9) & (0.0,0.8) & (--0.8,0.0) & (--2.5,--1.2)  \\
\hline
TPC track selection & 5.7 & 1.2 & 0.9 & 0.8 & 3.3\\
Particle identification &  -& - & 1.3 & 0.6 &- \\
Muon tracking efficiency &2.0 & 2.0 & - & - & 3.0 \\
Muon matching efficiency & 0.5 & 0.5 & - & - & 0.5 \\
Trigger efficiency & 2.0 & 2.0 & $^{+5.9}_{-10.1}$ & $^{+5.9}_{-10.1}$ & 1.6  \\
V0C Trigger efficiency & - & - & - & - & 3.4 \\
Offline V0 veto efficiency & 2.7 & 3.5 & 2.1 & 2.1 &1.2 \\
Signal extraction & 3.6 & 2.2 & 2.1  & 1.9 & 3.0 \\
Feed-down & $^{+0.0}_{-1.3}$ & 1.0 & $^{+0.6}_{-1.0}$& $^{+1.0}_{-0.6}$& $^{+0.0}_{-1.4}$ \\
Luminosity uncorrelated&3.3 & 3.3 & 2.3 & 2.3 & 3.0 \\
Luminosity correlated& 1.6&1.6&1.6&1.6&1.6 \\
Branching ratio \cite{Agashe:2014kda} & 0.6 & 0.6 & 0.4 & 0.4 & 0.6 \\
\hline
Total & 8.7 & 6.4 & 11.1 & 11.0 & 7.7\\
\hline
\end{tabular}
\end{center}
\caption{Summary of the contributions to the systematic uncertainty, in percent, for the  $\jpsi$ cross section in the different rapidity intervals.}
\label{Tab:SU}
\end{table}

\section{Results}
\subsection{Cross sections for exclusive production of \jpsi in p--Pb\ collisions}

The measured cross section for the exclusive production of \jpsi in collisions of protons with lead nuclei is computed according to
\begin{equation}
\frac{{\rm d}\sigma}{{\rm d}y} =\frac{N_{\jpsi}}{\eXa \cdot \mathrm{BR}(\jpsi\to l^+l^-)\cdot
\mathcal{L}_{\mathrm{int}} \cdot\Delta y}\,
\label{Eq:XS}
\end{equation}
where $\Delta y$ represents the width of the rapidity region where the measurement is performed and the branching ratios are  $\mathrm{BR}(\jpsi\to \mathrm{e}^+\mathrm{e}^-) = (5.97 \pm 0.03$)\% and  $\mathrm{BR}(\jpsi\to \mu^+\mu^-) = (5.96\pm0.03)$\% according to ~\cite{Agashe:2014kda}. The values that\ go into this equation are listed in Table~\ref{Tab:Inputs}, while the measured cross sections are given  in Table~\ref{Tab:XS}.

For the mid-rapidity analysis the measurements in the muon and electron channels are computed independently. The cross sections  in the rapidity range $-0.8<y<0$ ($0<y<0.8$) are 10.7$\pm$1.3 (11.1$\pm$1.8) $\mu$b for the electron channel and  9.6$\pm$1.0 (10.4$\pm$1.0) $\mu$b for the muon channel, where the errors represent the statistical uncertainty. These measurements have been averaged, weighting each cross section with its statistical uncertainty.

\subsection{Cross sections for exclusive photoproduction of \jpsi off protons}

The relation between ${\rm d}\sigma/{\rm d}y$ and the cross section for the photoproduction  of \jpsi off protons,  $\sigma (\gamma + {\rm p} \rightarrow \jpsi + {\rm p})$, is given by
\begin{equation}
\frac{{\rm d}\sigma}{{\rm d}y} = k \frac{\mathrm{d}n}{\mathrm{d}k}\sigma (\gamma + {\rm p} \rightarrow \jpsi + {\rm p})\,
\label{Eq:gpXS}
\end{equation}
where $k=0.5{\times}M_{\jpsi}\exp{(-y)}$ is the photon energy in the laboratory frame and $k\mathrm{d}n/\mathrm{d}k$ the flux of photons with energy $k$ emitted by the lead nucleus. Using  STARLIGHT, the flux has been computed in impact parameter space and convoluted with the probability of no hadronic interaction. The uncertainty in the flux is obtained varying the radius of the lead nucleus, used in the nuclear form factor, by $\pm0.5$  fm, which corresponds to the nuclear skin thickness.

The measured cross sections are given in Table~\ref{Tab:XS}, and also include the value of the centre-of mass energy of the photon-proton system, $\langle W_{\gamma {\rm p}} \rangle$, computed as the average of $W_{\gamma {\rm p}}$ over the rapidity interval  weighted by the photoproduction cross section predicted by  STARLIGHT.

\begin{table}

\begin{center}
\small

\begin{tabular}{cccccc}
\hline
 Rapidity &  $\frac{{\rm d}\sigma}{{\rm d}y}$\ ($\mu$b) & $k \frac{\mathrm{d}n}{\mathrm{d}k}$  &   $W_{\gamma {\rm p}}$ (GeV) & $\langle W_{\gamma {\rm p}} \rangle$ (GeV)
&$\sigma (\gamma + {\rm p} \rightarrow \jpsi + {\rm p}) (\mathrm{nb}) $\\ \hline
(1.9, 2.7)     &  6.9 $\pm$ 1.4  $\pm$ 0.6  & 163.6$\pm$1.5 & (40.8, 60.9)   & 50.4  & 42  $\pm$ 9  $\pm$ 4  $\pm$ 1   \\
(1.2, 1.9)     & 8.7  $\pm$ 1.5  $\pm$ 0.6  & 140.2$\pm$1.5 & (60.9, 86.4)   & 73.1  & 62  $\pm$ 11 $\pm$ 4  $\pm$ 1   \\
(0.0, 0.8)     & 10.6 $\pm$ 0.8  $\pm$ 1.2  & 104.3$\pm$1.5 & (105.5, 157.4) & 129.9 & 101 $\pm$ 8  $\pm$ 11 $\pm$ 1   \\
(--0.8, 0.0)   & 10.0 $\pm$ 0.8  $\pm$ 1.1  &  79.4$\pm$1.5 & (157.4, 234.9) & 193.3 & 126 $\pm$ 10 $\pm$ 14 $\pm$ 2   \\
(--2.5, --1.2) &  7.1 $\pm$ 1.0  $\pm$ 0.5  &  36.5$\pm$1.4 & (286.9, 549.5) & 391.2 & 194 $\pm$ 27 $\pm$ 15 $\pm$ 7   \\
\hline
\end{tabular}
\end{center}
\caption{Measured differential cross sections ${\rm d}\sigma/{\rm d}y$ for  exclusive $\jpsi$ photoproduction off protons in ultra-peripheral p--Pb (positive rapidity values) and Pb--p (negative rapidity values) collisions at $\sqrt{s_{\rm NN}}=5.02$ TeV in the different rapidity intervals.  Values of the photon flux $k \mathrm{d}n/\mathrm{d}k$, the interval in \Wgp corresponding to the rapidity, the average \Wgp ($\langle W_{\gamma {\rm p}} \rangle$) and the extracted cross section $\sigma (\gamma + {\rm p} \rightarrow \jpsi + {\rm p})$. The first uncertainty in the cross sections is statistical, the second is\ systematic and the third comes from the uncertainty in the photon flux.}
\label{Tab:XS}
\end{table}

\begin{figure}[t]
\begin{center}$
\begin{array}{c}
\includegraphics[width=0.98\textwidth]{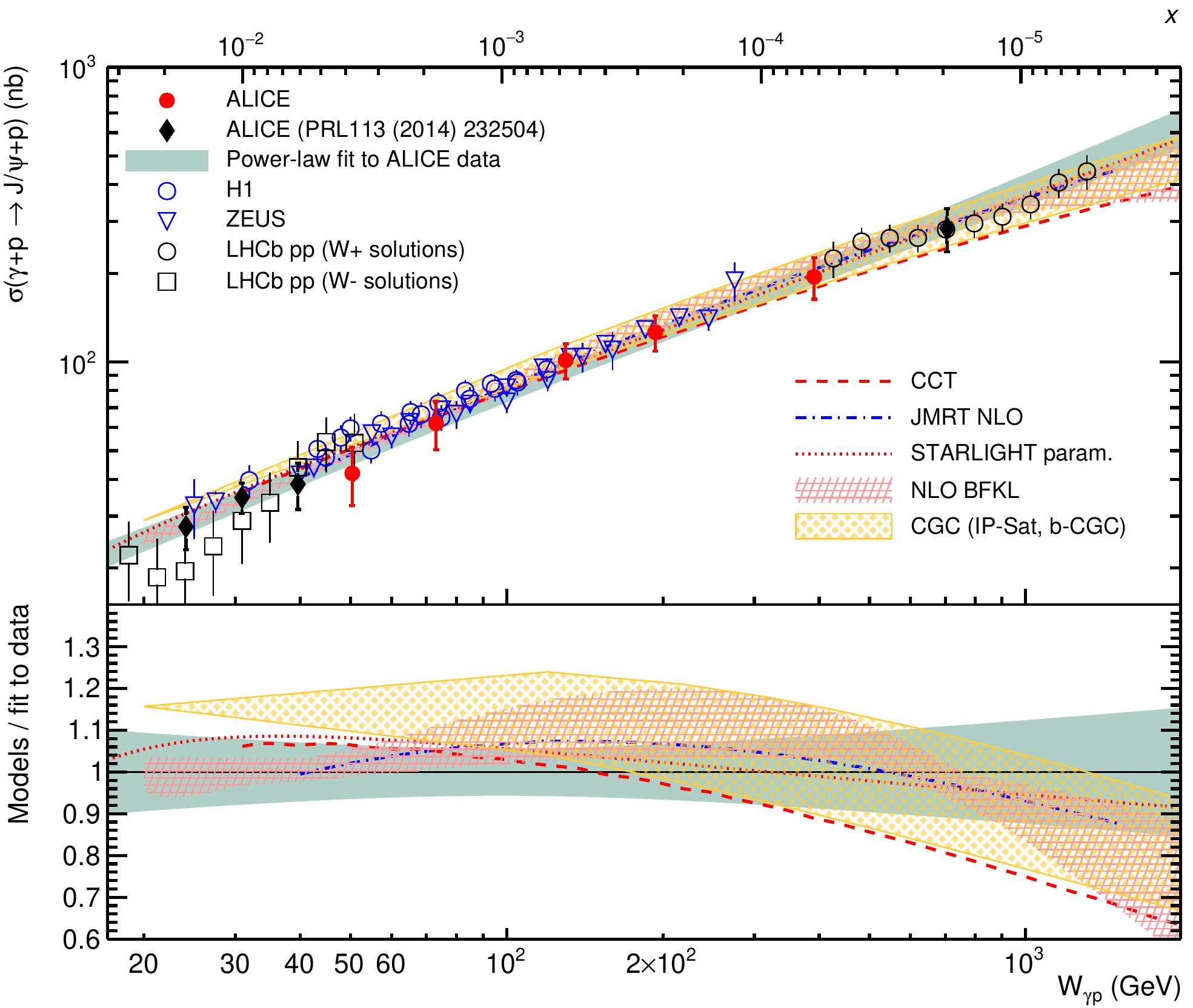}
\end{array} $
\end{center}
\caption{(Upper panel) ALICE data (red symbols) on exclusive photoproduction of \jpsi off protons as a function of the centre-of-mass energy of the photon--proton system \Wgp, obtained in collisions of protons and lead nuclei at $\snn=5.02$ TeV, including results from~\cite{TheALICE:2014dwa},  compared to a power-law fit, to data from HERA\cite{Chekanov:2002xi,Alexa:2013xxa}, to the solutions from LHCb\cite{Aaij:2013jxj}\ and to theoretical models (see text). The uncertainties are the quadratic sum of the statistical and systematic uncertainties. (Lower panel) Ratio of the models shown in the upper panel\ to the power law fit through the ALICE data points. The Bjorken x value corresponding to \Wgp is also displayed on the top of the Figure, see text for details.}
\label{fig:Wdep}
\end{figure}	

\subsection{Energy dependence of the exclusive photoproduction of \jpsi off protons}

Figure~\ref{fig:Wdep} shows all ALICE measurements, including those from~\cite{TheALICE:2014dwa}, for the exclusive photoproduction of \jpsi off protons as a function of $\Wgp$. The data cover the $\Wgp$ range from 24 to 706 GeV, which corresponds roughly to three orders of magnitude in  $x$, from $\sim 2\times 10^{-2}$ to $\sim2\times 10^{-5}$. The error bars are the quadratic sum of the statistical and the total systematic uncertainties.

The same figure shows the result of a $\chi^2$-fit of a power law $N(\Wgp/W_0)^\delta$ to the full set of ALICE data, with $W_0=90.0$ GeV as was done before in HERA analyses~\cite{Newman:2013ada}.\ The fit takes into account the statistical and systematic uncertainties, according to the technique used by H1 in  \cite{Aaron:2009ab}. The parameters found by the fit are\  $N=71.8\pm4.1$ nb and $\delta=0.70\pm0.05$ with a  correlation of $-0.06$ between both parameters. The quality of the fit is $\chi^2 = 1.21$ for 7 degrees of freedom.
The value of the exponent is compatible to that found using previous ALICE data~\cite{TheALICE:2014dwa}, as well as with that found by HERA experiments~\cite{Chekanov:2002xi,Alexa:2013xxa}.

The comparison of ALICE measurements to data from other experiments as well as to the results from different models is also shown in Fig.~\ref{fig:Wdep}.  HERA~\cite{Chekanov:2002xi,Alexa:2013xxa} and ALICE data are compatible within uncertainties. LHCb measured the exclusive production of \jpsi in pp collisions, where the photon source can not be identified. Thus the extraction of the photoproduction cross section is not possible without further assumptions. For each measurement they reported two solutions~\cite{Aaij:2014iea} which also agree with ALICE measurements.

ALICE measurements are also compared to theory in Fig.~\ref{fig:Wdep}. The JMRT group~\cite{Jones:2013pga} has two computations, one is based on the leading-order (LO) result from~\cite{Ryskin:1992ui} with the addition of some corrections to the cross section, while the second includes also the main contributions expected from a next-to-leading order (NLO) result. The parameters of both models have been obtained by a fit to the same data and their energy dependence is rather similar, so only the NLO version is shown.
Recently, three new studies have appeared, describing the $W({\gamma}p)$ dependence of the exclusive \jpsi cross section  in terms of a colour dipole model\ \cite{Armesto:2014sma} (CGC) or of the BFKL evolution of HERA values (HERA Fit 2) with a photoproduction scale $M^{2} = 2.39$\ $\mathrm{GeV}^{2}$ \cite{PhysRevD.94.054002} (NLO(BFKL)). These are shown as bands in the figure. A third model, based on the colour dipole approach, and incorporating the energy dependence of geometrical fluctuations of the proton structure in the impact parameter plane \cite{Cepila:2016uku} is also shown (CCT).  The models are in reasonable agreement with our data.
Finally the STARLIGHT parameterisation relies on a power-law fit to  fixed-target and HERA data. This model also agrees with our measurement.

\section{Summary}

The ALICE Collaboration has measured the photoproduction of \jpsi mesons off protons in \pPb interactions. New measurements, summarised in Table~\ref{Tab:XS}, at central, semi-backward and semi-forward\ rapidities\ are added to those previously given at forward and backward rapidities. Each rapidity interval corresponds to a given energy for exclusive photoproduction in photon--proton interactions. The data agree with the previous ALICE measurements at forward and backward\ rapidities, with  the LHCb results in pp interactions and with previous HERA measurements over a smaller energy range. The ALICE measurements\ are consistent with a power law dependence $\sigma(\gamma p \rightarrow \jpsi p) \sim W_{\gamma p}^{\delta}$, with $\delta = 0.70 \pm 0.05.$ Several models, based on different physics assumptions, reproduced this behaviour within the current
 experimental uncertainties. The model predictions around and above 1 TeV are not very precise although all tend to go below the extrapolation of the fit, as shown in the lower panel of Fig.~\ref{fig:Wdep}. This energy range
 will be reachable with the new LHC data from Run 2 and the data to be collected in Run 3 and Run 4. The data presented here, augmented with the results from future measurements, will be a powerful tool to better understand the role of saturation at the highest energies.

\bibliographystyle{utphys}
\bibliography{CandSF}

\newenvironment{acknowledgement}{\relax}{\relax}
\begin{acknowledgement}
\section*{Acknowledgements}

The ALICE Collaboration would like to thank all its engineers and technicians for their invaluable contributions to the construction of the experiment and the CERN accelerator teams for the outstanding performance of the LHC complex.
The ALICE Collaboration gratefully acknowledges the resources and support provided by all Grid centres and the Worldwide LHC Computing Grid (WLCG) collaboration.
The ALICE Collaboration acknowledges the following funding agencies for their support in building and running the ALICE detector:
A. I. Alikhanyan National Science Laboratory (Yerevan Physics Institute) Foundation (ANSL), State Committee of Science and World Federation of Scientists (WFS), Armenia;
Austrian Academy of Sciences and Nationalstiftung f\"{u}r Forschung, Technologie und Entwicklung, Austria;
Ministry of Communications and High Technologies, National Nuclear Research Center, Azerbaijan;
Conselho Nacional de Desenvolvimento Cient\'{\i}fico e Tecnol\'{o}gico (CNPq), Universidade Federal do Rio Grande do Sul (UFRGS), Financiadora de Estudos e Projetos (Finep) and Funda\c{c}\~{a}o de Amparo \`{a} Pesquisa do Estado de S\~{a}o Paulo (FAPESP), Brazil;
Ministry of Science \& Technology of China (MSTC), National Natural Science Foundation of China (NSFC) and Ministry of Education of China (MOEC) , China;
Ministry of Science and Education, Croatia;
Centro de Aplicaciones Tecnol\'{o}gicas y Desarrollo Nuclear (CEADEN), Cubaenerg\'{\i}a, Cuba;
Ministry of Education, Youth and Sports of the Czech Republic, Czech Republic;
The Danish Council for Independent Research | Natural Sciences, the Carlsberg Foundation and Danish National Research Foundation (DNRF), Denmark;
Helsinki Institute of Physics (HIP), Finland;
Commissariat \`{a} l'Energie Atomique (CEA) and Institut National de Physique Nucl\'{e}aire et de Physique des Particules (IN2P3) and Centre National de la Recherche Scientifique (CNRS), France;
Bundesministerium f\"{u}r Bildung, Wissenschaft, Forschung und Technologie (BMBF) and GSI Helmholtzzentrum f\"{u}r Schwerionenforschung GmbH, Germany;
General Secretariat for Research and Technology, Ministry of Education, Research and Religions, Greece;
National Research, Development and Innovation Office, Hungary;
Department of Atomic Energy Government of India (DAE), Department of Science and Technology, Government of India (DST), University Grants Commission, Government of India (UGC) and Council of Scientific and Industrial Research (CSIR), India;
Indonesian Institute of Science, Indonesia;
Centro Fermi - Museo Storico della Fisica e Centro Studi e Ricerche Enrico Fermi and Istituto Nazionale di Fisica Nucleare (INFN), Italy;
Institute for Innovative Science and Technology , Nagasaki Institute of Applied Science (IIST), Japan Society for the Promotion of Science (JSPS) KAKENHI and Japanese Ministry of Education, Culture, Sports, Science and Technology (MEXT), Japan;
Consejo Nacional de Ciencia (CONACYT) y Tecnolog\'{i}a, through Fondo de Cooperaci\'{o}n Internacional en Ciencia y Tecnolog\'{i}a (FONCICYT) and Direcci\'{o}n General de Asuntos del Personal Academico (DGAPA), Mexico;
Nederlandse Organisatie voor Wetenschappelijk Onderzoek (NWO), Netherlands;
The Research Council of Norway, Norway;
Commission on Science and Technology for Sustainable Development in the South (COMSATS), Pakistan;
Pontificia Universidad Cat\'{o}lica del Per\'{u}, Peru;
Ministry of Science and Higher Education and National Science Centre, Poland;
Korea Institute of Science and Technology Information and National Research Foundation of Korea (NRF), Republic of Korea;
Ministry of Education and Scientific Research, Institute of Atomic Physics and Romanian National Agency for Science, Technology and Innovation, Romania;
Joint Institute for Nuclear Research (JINR), Ministry of Education and Science of the Russian Federation and National Research Centre Kurchatov Institute, Russia;
Ministry of Education, Science, Research and Sport of the Slovak Republic, Slovakia;
National Research Foundation of South Africa, South Africa;
Swedish Research Council (VR) and Knut \& Alice Wallenberg Foundation (KAW), Sweden;
European Organization for Nuclear Research, Switzerland;
National Science and Technology Development Agency (NSDTA), Suranaree University of Technology (SUT) and Office of the Higher Education Commission under NRU project of Thailand, Thailand;
Turkish Atomic Energy Agency (TAEK), Turkey;
National Academy of  Sciences of Ukraine, Ukraine;
Science and Technology Facilities Council (STFC), United Kingdom;
National Science Foundation of the United States of America (NSF) and United States Department of Energy, Office of Nuclear Physics (DOE NP), United States of America.    
\end{acknowledgement}


\newpage
\appendix
\section{The ALICE Collaboration}
\label{app:collab}
\begin{flushleft} 
\bigskip 

S.~Acharya$^{\rm 139}$, 
F.T.-.~Acosta$^{\rm 20}$, 
J.~Adam$^{\rm 37}$, 
D.~Adamov\'{a}$^{\rm 93}$, 
A.~Adler$^{\rm 74}$, 
J.~Adolfsson$^{\rm 80}$, 
M.M.~Aggarwal$^{\rm 98}$, 
G.~Aglieri Rinella$^{\rm 34}$, 
M.~Agnello$^{\rm 31}$, 
N.~Agrawal$^{\rm 48}$, 
Z.~Ahammed$^{\rm 139}$
% ...

\bigskip

\bigskip 

\textbf{\Large Affiliation Notes}

\bigskip 

% ...
$^{\rm V}$ Also at: Institute of Theoretical Physics, University of Wroclaw, Poland\\

\bigskip

\bigskip 

\textbf{\Large Collaboration Institutes}

\bigskip 

% ...
$^{20}$ Department of Physics, University of California, Berkeley, California, United States\\
% ...
$^{31}$ Dipartimento DISAT del Politecnico and Sezione INFN, Turin, Italy\\
% ...
$^{34}$ European Organization for Nuclear Research (CERN), Geneva, Switzerland\\
% ...
$^{37}$ Faculty of Nuclear Sciences and Physical Engineering, Czech Technical University in Prague, Prague, Czech Republic\\
% ...
$^{48}$ Indian Institute of Technology Bombay (IIT), Mumbai, India\\
% ...
$^{74}$ Johann-Wolfgang-Goethe Universit\"{a}t Frankfurt Institut f\"{u}r Informatik, Fachbereich Informatik und Mathematik, Frankfurt, Germany\\
% ...
$^{80}$ Lund University Department of Physics, Division of Particle Physics, Lund, Sweden\\
% ...
$^{93}$ Nuclear Physics Institute of the Czech Academy of Sciences, \v{R}e\v{z} u Prahy, Czech Republic\\
% ...
$^{98}$ Physics Department, Panjab University, Chandigarh, India\\
% ...
$^{139}$ Variable Energy Cyclotron Centre, Homi Bhabha National Institute, Kolkata, India\\
% ...

\bigskip 

\end{flushleft}  
\end{document}